\documentclass[a4paper]{article}
\usepackage[english]{babel} \usepackage{hyperref} \usepackage{float}
\usepackage[utf8]{inputenc} \usepackage{amsmath} \usepackage{graphicx}
\usepackage[colorinlistoftodos]{todonotes}\usepackage{tikz}
\usepackage{pdfpages} \usepackage{listings}
\usepackage{cite}
\usetikzlibrary{arrows,positioning,shapes.geometric}
\usetikzlibrary{arrows,positioning,shapes.geometric}
\title{\textbf{Is there a chiral dark dynamo in the universe induced by quantum correction, Nieh-Yan 
gravity and Barbero-Immirzi field?}}
\date{}
\begin{document}
\maketitle
\vspace{0.001cm} 

\begin{center}
{\large Zhi Fu Gao\footnote{1. Xinjiang Astronomical Observatory, Chinese Academy of Sciences, Urumqi, Xinjiang, 830011, 
China, zhifugao@xao.ac.cn;}}  
\end{center}
\begin{center}
{\large Biao Peng Li\footnotemark[1]\footnote{1. Xinjiang Astronomical Observatory, Chinese Academy of Sciences, Urumqi, 
Xinjiang, 830011, China; 2. University of Chinese Academy of Sciences, Beijing, China;}}
\end{center}
\begin{center}
{\large L.C. Garcia de Andrade\footnote{Cosmology and gravitation group, Departamento de
F\'{\i}sica Te\'{o}rica - IF - UERJ - Rua S\~{a}o Francisco Xavier
524, Rio de Janeiro, RJ, Maracan\~{a}, CEP:20550. Institute for Cosmology and philosophy of Nature, 
 Croatia. e-mail:luizandra795@gmail.com}}
\end{center}
\vspace{0.001cm}
\begin{abstract} 
\vspace{0.1cm}
Previously, Bombagcino et al.\,(2021) investigated the role of Immirzi parameter when promoted to a field in 
Einstein-Cartan-Holst black hole. In this framework they found that the Immirzi field acts similar to the axion field, 
as both axial pseudo-vector and vectorial torsion trace appear to be expressed in terms of the 4-gradient of the Immirzi 
parameter. In this paper we introduced two important ingredients absent in the previous work: the torsion mass, 
significant for the torsion detection the Large Hadron Collider (LHC), and the quantum correction proportional to the 
4-divergent of torsion squared. Without the quantum correction, a simple analytical solution is obtained, while the more 
complicated field equations incorporating the BI field are obtained also analytically. The  lower bound of quantum 
correction parameter is determined in terms of the torsion trace mass squared and axial torsion squared. Our findings 
reveal that in the late universe, the BI parameter approaches infinity restoring to the Einstein-Cartan theory in the 
early universe with the dynamical reduction of the Immirzi parameter to a constant BI parameter. Additionally, we 
derive analytical solutions for magnetic dynamos in the early universe, demonstrating that magnetic helicity is 
proportional to chiral chemical potential. A magnetic field at the QCD phase is found out of $10^{17}$\,G, without 
quantum correction. Furthermore, from this dark magnetogenesis, we estimate light torsion with mass of the order of 
$1$\,TeV, An example of unitary preserved Lagrangian with axion as an Immirzi field is obtained. In the present 
universe we find a magnetic field strength of approximately $10^{-12}$\,G which is quite close to the range found 
by Miniati et al (Phys Rev Lett 121: 021301, 2018) at the QCD threshold, between $10^{-18}-10^{-15}$\,G. Given that 
unitary violation on theoretical grounds may indicate new physics, exploring unitary violations in dark magnetogenesis 
could be particularly intriguing.
\end{abstract}

\section{Introduction}\label{sec1}
Recently, several authors\,\cite{1,2,3,4} have promoted the Barbero-Immirzi\,(BI) parameter in Loop Quantum Gravity\,(LQG) 
to a dynamical field, reducing it to a scalar field. More recently Bombacigno et al.\,\cite{5} have investigated the 
inflationary cosmology of Bianchi type I models in the Einstein-Cartan-Holst framework, with Nieh-Yan\,(YN) term. They 
explored the NY topological invariant in the context of projectively invariant models, specifically focusing on the 
Bianchi I cosmological model. Some authors investigated black hole hair through a scalar field\,\cite{6,7}. Interestingly, 
in their massless torsion theory of gravity, they demonstrated that the types of torsion—axial pseudo, axial pseudo-vector, 
and vectorial torsion traces—could be expressed in terms of the 4-gradient of an Immirzi dynamical field ${\gamma}(x)$. 
This is analogous analogy to the work of Duncan et al.(1992)\,\cite{8} on the transmutation of torsion into an action to 
obtain magnetic fields as black hole hair. 

The quadratic Poincaré Gauge theory of gravity's stable pseudo-scalar degree of freedom is a suitable dark matter candidate 
\cite{9}. Einstein-Cartan gravity inherently includes four-fermion and scalar-fermion interactions due to torsion associated 
with spin. These interactions present a novel mechanism for generating singlet fermions in the Early Universe, which can act 
as dark matter particles \cite{10}. Shapiro (2002)\,\cite{11} have reviewed many quantum aspects of torsion theory and 
discussed the possibility of the space-time torsion to exist and to be detected. In the EC gravity, there exists geometric 
quantities other than the Ricci curvature, such as the YN term and the Holst term \cite{12,13,14,15}, owing to the presence 
of the torsion. As shown in He et al.(2024) \cite{16}, the scalaron becomes dynamical by allowing the NY and/or the Holst 
terms together with the Ricci scalar to be present with general combinations up to their quadratic order. In this work, we 
present Riemann flat theory of gravity based on Einstein-Cartan-Holst-Nieh-Yan theory with presence of quantum correction term proportional to the covariant derivative of axial 
torsion squared. In our framework, the Immirzi field functions as an axion scalar field. It is shown that in the 
early universe, the Immirzi field can be considered as the usual BI constant numerical parameter. However, in the late 
universe, this dynamical Immirzi field reduces to zero, indicating that the ECHNY theory reverts to EC gravity over time.

Very recently, Garcia\,\cite{17} investigated the effects of BI and quantum corrections on EC gravity with 
very light inflatons, and provided an important analytical solution for chiral dynamos in the early universe.
The author explored dark torsion oscillations and bounces, providing insights into the interplay between torsion, dark 
photons, and axions in cosmological contexts. Meanwhile, this study addresses a fundamental question: Would 
chiral dynamos exist as axionic dark dynamos in the early universe? 

Additionally, Miniati et al.\,\cite{18} explored axionic dynamos at the QCD threshold, finding a magnetic 
field of $B=10^{-13}$\,G on a characteristic scale of 20 parsecs. This study complements Garcia’s work by demonstrating 
another context in which axionic mechanisms influence cosmic magnetic fields. Despite finding weaker magnetic fields, 
this research provides valuable insights into the generation of cosmological magnetic fields through axion dynamics.

In contrast, Berera et al.\,\cite{19}, in their 1999 study, investigated cosmic magnetic fields generated during the warm 
inflationary phase within the framework of the Grand Unified Theory (GUT). They proposed a scenario that yields 
significantly stronger magnetic fields, approximately $B_{\mathrm{GUT}}\sim10^{45}$\,G. This substantial difference in 
field strength highlights the diverse mechanisms through which magnetic fields can be produced in the early universe.

By linking these studies, we gain a comprehensive understanding of the various theoretical models that predict the 
generation of magnetic fields in the early universe, ranging from the dark torsion oscillations in \cite{17} 
to the axionic dynamos in \cite{18} and the GUT-based scenario in \cite{19}. The reminder of this paper is organized as 
follows. In Section\,\ref{sec2}, we will derive and solve the axion magnetic field wave equation analytically. This raises 
the question: Could this axion chiral dark dynamo involve the Immirzi parameter? Section\,\ref{sec3} explores the field 
equations of Holst NY gravity without magnetic fields. Section\,\ref{sec4} addresses the restoration of unitarity in 
one-loop corrections with torsion terms up to fourth-order. Finally, Section\,\ref{sec5} is left for summary and outlook.

\section{Immirzi Dynamical Field and Dark Magnetogenesis in Cartan-Holst-Nieh-Yan Portal via Unitary Violation}\label{sec2}
In their 1994 study, Carroll and Field \cite{19} explored the consequences of propagating torsion in connection-dynamic 
theories of gravity, highlighting the constraints and implications of such theories. Their work provided insights into the 
behavior of torsion degrees of freedom and their interactions with matter fields, which is fundamental for the framework 
discussed in this section.

More recently, Tukhashvili and Steinhardt \cite{20} investigated cosmological bounces induced by a fermion condensate, 
demonstrating how nonsingular bounces can occur and the stability of these scenarios in the early universe. These findings 
introduce mechanisms for cosmological bounces that are relevant to our discussion of dark magnetogenesis and unitary 
violation.

In this section, we will present a framework of the Cartan-Holst-Nieh-Yan (CHNY) action, where the Immirzi parameter is promoted to a dynamical field 
${\gamma}(x)$. Besides the torsion mass, we also add a new interesting term to the action: the $b$ parameter in front of 
a divergence of axial torsion pseudo-vector $S$ squared. Here, $S$ represents, in compact notation, the four-dimensional 
vector. The main Lagrangian is 
given by 
\begin{equation}
{\cal{L}}_{\mathrm{BI}/\mathrm{CHNY}}\supset{b{I_{\mathrm{NY}}}^{2}+\frac{1}{2}[M_{s}^{2}S^{2}+M_{T}^{2}T^{2}]+{\gamma}(x)[I_{\mathrm{NY}}+{\cal{H}}]} .
\label{1}
\end{equation}
Here, $bI_{\mathrm{NY}}^{2}$ represents the quantum correction term, with $I_{\mathrm{NY}}$ being the NY invariant\,\footnote{$I_{\mathrm{NY}}$ is a 
topological term that appears in the context of theories involving torsion, such as Einstein-Cartan theory. It can be defined as the 
divergence of the axial torsion pseudo-vector $S$, i.e. $I_{\mathrm{NY}}= {\nabla}_{i}S^{i}$\,\cite{17}.}. 
$\frac{1}{2}[M_{s}^{2}S^{2}+M_{T}^{2}T^{2}]$ includes the mass terms for the torsion vector $S$ and its trace $T$, with $M_{S}$ and 
$M_{T}$ being their respective masses. $\gamma(x)[I_{\mathrm{NY}}+{\cal{H}}]$ is the coupling between the Immirzi field $\gamma(x)$ 
and the sum of $I_{\mathrm{NY}}$ and the Holst term $\cal{H}$\,\cite{21}. We omit the name ``Einstein" in the Cartan-Holst-Nieh-Yan Lagrangian 
because we consider the Riemannian metric to be that of Minkowski flat spacetime. This choice aligns with the experiments 
and torsion data from the LHC at CERN (European Organization for Nuclear Research).

In a curved spacetime with metric determinant $g$ the covariant divergence of $S$ can be 
expressed as ${\partial}_{i}(\sqrt{-g}S^{i})$, then we have
\begin{equation}
I_{\mathrm{NY}}= {\nabla}_{i}S^{i}= {\partial}_{i}(\sqrt{-g}S^{i}) ,
\label{2}
\end{equation}
where $i,j=0, 1, 2, 3$. If we take the variation $\cal{L}_{\mathrm{BI}/\mathrm{CHNY}}$ with respect to $T$  in the presence of the 
Immirzi field $\gamma(x)$ that couples to the Nieh-Yan invariant and Holst term in the action, leading to an algebraic 
expression for $T$ as  
\begin{equation}
T= -\frac{{\gamma}(x)}{M^{2}_{T}}S ,
\label{3}
\end{equation}
where ${\gamma(x) \cal{H} }= \gamma(x)T\cdot S$ (${\cal{H}}\supset T\cdot S$) is used.
Taking the variation of the above Lagrangian with respect to the torsion vector $S$, we have
\begin{equation}
\frac{\partial \cal{L}_{\mathrm{BI}/\mathrm{CHNY}}}{\partial S}=2bI_{\mathrm{NY}}\frac{\partial I_{\mathrm{NY}} }{\partial S}+M_{s}^{2}S+\gamma(x)\frac{\partial I_{\mathrm{NY}} }
{\partial S} .
\label{4}
\end{equation}
Since $I_{\mathrm{NY}}= {\nabla}_{i}S^{i}$, then $\frac{\partial I_{\mathrm{NY}} }{\partial S}=\nabla_{i}$. To derive the equation of motion, 
we consider the Euler-Lagrange equation
\begin{equation}
\frac{\partial \cal{L}_{\mathrm{BI}/\mathrm{CHNY}}}{\partial S}-\partial_{\mu}(\frac{\partial \cal{L}_{\mathrm{BI}/\mathrm{CHNY}}}{\partial (\partial_{\mu}S)})=0.
\label{5}
\end{equation}
Here, the partial derivative ${\partial}$ is taken with respect to cosmic time $t$.
Simplifying this expression yields
\begin{equation}
2b{\partial}^{2}S+{\dot{\gamma}}-2S\frac{{\gamma}^{2}}{M^{2}_{T}}=0 .
\label{6}
\end{equation}
From this differential equation, we notice that in the absence of quantum correction ($b=0$) it simplifies to a differential 
equation for the homogeneous $\gamma$ function, However, to explore the more general case, we retain the $b$ term.

Next, varying the Lagrangian with respect to the Immirzi field, and considering the Euler-Lagrange equation, we get
\begin{equation}
{\partial}^{2}S+\frac{b}{M^{2}_{T}}[{\dot{\gamma}}S^{2}-2S\dot{S}]=0 ,
\label{7}
\end{equation}
where $\dot{S}$ represents the time derivative of $S$. This equation describes the coupled dynamics of the torsion vector 
$S$ and the Immirzi field $\gamma$, considering both the quantum correction and the mass term. 
Assuming $S$ is constant, we find an approximation relation,
\begin{equation}
\frac{\dot{\gamma}}{\gamma}\approx{\frac{S}{M^{2}_{T}-2bS^{2}}} .
\label{8}
\end{equation}
Note that before solving Equation (\ref{9}), an interesting physical determination of the quantum correction bound of 
parameter $b$, can be obtained as follows: We know that from  Barman et al \cite{22} and \cite{23}, the torsion mass is 
around $1.7$\,TeV, which is significantly larger than the axial torsion $S$. Therefore,
the denominator in the right-hand-side of the last equation must be positive, which yields an upper bound 
for the quantum correction parameter $b$ as
\begin{equation}
\frac{M^{2}_{T}}{2S^{2}}\ge{b} .
\label{9}
\end{equation}
Now by solving Equation (\ref{8}), one shows that 
\begin{equation}
{\gamma}={\gamma}_{0}\exp[\frac{S}{M^{2}_{T}-2bS^{2}}t] .
\label{10}
\end{equation}
At $t=0$, we note that ${\gamma}_{0}$ is the original BI constant numerical expression. Here, ${\gamma}$ is a free parameter.
As time progresses, the Immirzi parameter  ${\gamma}$ evolves exponentially, influenced by the torsion vector $S$, torsion 
mass $M_{T}$ and quantum correction parameter $b$. $M^{2}_{T}-2bS^{2}$ must be positive, ensuring stable evolution of 
$\gamma$. This equation suggests a dynamic Immirzi parameter, impacting quantization of spacetime and behavior in LQG.
We also note that, from the solution for the Immirzi parameter in the early universe the Immirzi dynamical parameter reduces 
to the original LQG BI numerical parameter that can be determined by Panza et al \cite{24} at TeV scales of 
particle physics. In the next section we will investigate the chiral dynamos in the CHNY gravity.

\section{Dark chiral magnetogenesis via magnetic helicity and Immirzi parameter in Cartan--Holst-Nieh-Yan model}
\label{sec3} 
The Nieh-Yan modified teleparallel gravity includes an additional term called the NY term. This term is introduced into the gravitational 
action and it breaks the parity symmetry in gravity\,\cite{25,26}.The Nieh-Yan invariant is a topological quantity involving 
torsion, which has been used successfully in cosmological models\,\cite{25,26,27}. It helps describe certain properties of 
the universe in the context of teleparallel gravity and its modifications.

Here we will explore a fascinating problem originally investigated by Carroll and Field \cite{28}, which involves 
the generation of primordial magnetic fields from primordial helicity. This approach can be applied to chiral dynamos 
studied in general relativity by Schober et al. \cite{29}, helping to simplify the magnetic wave equations for generating 
chiral dynamos.

To explore dynamo action generated by quantum correction $b$ and torsion mass, we add two electromagnetic terms to the 
Lagrangian from the previous section. First, we include the regular electromagnetic (EM) Maxwell term:
\begin{equation} 
F_{ij}F^{ij}=4F^{2} ,
\label{11}
\end{equation}
where $F_{ij}= 2{\partial}_{[i}A_{j]}$ is the EM field tensor. 

The other electromagnetic term involves the use 
of the dual of $F$ term, ${\tilde{F}}^{ij}={\epsilon}^{ijkl}F_{kl}$, to compose the axion-like term, 
\begin{equation}
{\cal{L}}_{\mathrm{Dual}}=\gamma{F\tilde{F}}+J^{5}A ,
\label{12}
\end{equation}
where $A$ is the EM field one-form. By varying the whole integral
\begin{equation}
{\cal{L}}_{\mathrm{total}}={\cal{L}}_{\mathrm{CHNY}}+{\cal{L}}_{\mathrm{M}}+{\cal{L}}_{\mathrm{Dual}}.
\label{13}
\end{equation}
we obtain the following equations. First, the variation of $\gamma$ yields
\begin{equation}
ST+{\partial}_{i}S^{i}+ F{\tilde{F}}=0 .
\label{14}
\end{equation}
A simple algebra leads to the equation from the variation of $T$, which coincides with Equation (\ref{10}) above. 
Then the equation for the electromagnetic field in terms of the axion-Immirzi parameter is given as
\begin{equation} 
dF+\dot{\gamma}{\wedge}F=J_{5} .
\label{15}
\end{equation}
This expression is written using Cartan's differential forms. The wedge product $\dot{\gamma}{\wedge}F$ represents 
the exterior product of the time derivative of the Immirzi field with the electromagnetic field tensor. This 
combination incorporates the antisymmetric properties of the differential forms, which are essential in electromagnetism 
and general relativity. The chiral current $J^{5}$ is expressed as 
\begin{equation} 
{\partial}_{i}F^{ij}+({\partial}_{i}{\gamma})F^{ij}=J^{j}_{5} .
\label{16}
\end{equation}
Considering a homogeneous Immirzi field, in terms of the vectorial Maxwell notation, one has
\begin{equation} 
{\nabla}\cdot{\textbf{E}}={\rho}_{5},
\label{17}
\end{equation}
and the other equations arise from 
\begin{equation} 
{\partial}_{0}F^{0a}+{\partial}_{b}F^{ba}+({\partial}_{0}{\gamma})F^{oa}=J^{a}_{5} .
\label{18}
\end{equation}
In vector analysis notation, this yields
\begin{equation}
{\partial}_{t}\textbf{E}+{\nabla}{\times}{\textbf{B}}+(\dot{\gamma}){\textbf{B}}={\textbf{J}}_{5} .
\label{19}
\end{equation}
The chiral current is given by
\begin{equation}
\textbf{J}_{5}={\mu}_{5}\textbf{B}
\label{20}
\end{equation}
and the magnetic helicity definition is
\begin{equation}
{\nabla}{\times}\textbf{B}=\lambda\textbf{B} ,
\label{21}
\end{equation}
where ${\mu}_{5}$ is the chiral chemical potential, and $\lambda$ is the magnetic helicity parameter. These 
expressions allow us to analyze the evolution and behavior of magnetic fields in the presence of chirality. 
In this paper we consider the chemical chiral potential as homogeneous but very recently Anzuini and Maggi 
\cite{30} have addressed axionic QCD without torsion but with non-homogeneous as we consider here.   
Substituting the last two expressions into Equation (\ref{19}) yields
\begin{equation}
{{\partial}^{2}}_{t}\textbf{B}-(\dot{\gamma}){\partial}_{t}{\textbf{B}}+({\lambda}^{2}-{\mu}_{5}{\lambda})\textbf{B}=0 .
\label{22}
\end{equation}
This is the derived magnetic wave equation taking into account the axion-Immirzi field and the effects of the chiral 
chemical potential and magnetic helicity.

To solve this magnetic wave equation, we need a solution for the Immirzi field. In the Einstein-Cartan-Nieh-Yan
 theory of gravity, the Immirzi parameter $\gamma$ is usually associated with torsion and mass. In this case, we assume 
that the variation of the Immirzi field is inversely proportional to the ratio of a constant $M^{2}_{T}$ and another 
constant $S_{0}$. Considering this physical model, the Immirzi field is typically expressed as
\begin{equation}
{\gamma}(x)=\frac{M^{2}_{T}}{S_{0}}t^{-1} ,
\label{23}
\end{equation}
where $S_{0}$ is a system characteristic constant. It represents a specific property or parameter of the system being 
studied, which is related to the torsion and Immirzi field.

Taking the time derivative of Equation (23), we obtain
\begin{equation}
\dot{\gamma}(x)=-\frac{M^{2}_{T}}{S_{0}}t^{-2} .
\label{24}
\end{equation}
Substituting this expression into the wave equation (\ref{22}) yields
\begin{equation}
\partial^{2}_{t}\textbf{B}+(\frac{M^{2}_{T}}{S_{0}}t^{-2}){\partial}_{t}{\textbf{B}}+({\lambda}^{2}-{\mu}_{5}
{\lambda})\textbf{B}=0 ,
\label{25}
\end{equation}
where the term $\frac{M^{2}_{T}}{S_{0}t^{-2}}$ represents the contribution of the Immirzi field's time derivative to the 
wave equation, while $\lambda^{2}-{\mu}_{5}{\lambda}$ captures the influence of magnetic helicity and chiral chemical 
potential on the magnetic field. This equation describes the evolution of the magnetic field $B$ in the presence of the 
time-dependent Immirzi field. However, due to the lack of necessary initial conditions, including the magnetic field and 
its first derivative, it is not possible to provide an exact numerical solution to this equation.  Another difficulty 
arises from the estimation of the parameters in the equation. For example, different authors have different estimates for 
the mass of trace torsion, $M_{\mathrm{T}}$. Gacia estimated $M_{\mathrm{T}} \sim 1$ \,MeV\,\cite{3} from 
\begin{equation}
M^{2}_{\mathrm{T/QCD}}\sim B_{\mathrm{QCD}}\sim 10^{17}~~\mathrm{G},
\label{26}
\end{equation}
where $B_{\mathrm{QCD}}$ represents the strength of the cosmic magnetic field at quantum-chromodynamics\,(QCD) 
 threshold $t\sim 10^{-5}$\,s, and the units transformations $1 \mathrm{G}=10^{-20}\,\mathrm{GeV}^{2}$ and $1 \mathrm{Hz}= 10^{-21}$\,GeV are used. Recently, Mavromatos et al.\cite{31} discussed the role of torsion in string theory 
and its implications for quantum gravity and cosmological tension,  and obtained a massive torsion trace $M_{\mathrm{T}}\sim 1$\,TeV at the QCD threshold. 
  
\begin{figure}[H]
    \centering
    \includegraphics[width=0.9\linewidth]{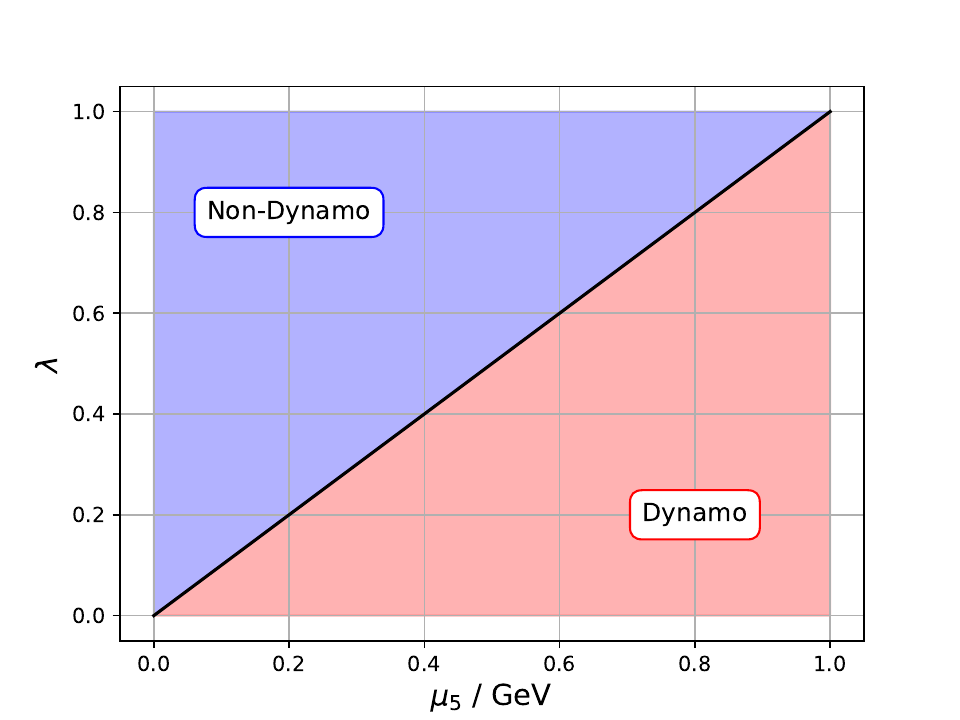}
    \caption{In this figure the blue part corresponds to magnetic helicity $\lambda \ge{{\mu}_{5}}$ or non-dynamo region.
The lower red part corresponds to ${\lambda}\le{{\mu}_{5}}$ representing the dynamo action region of the imbalance between 
the msgnetic helicity and chiral chemical potential.}
    \label{fig1}
\end{figure}

Now let's investigate Equation (\ref{25}). Given that the second derivative of the magnetic field drops too rapidly, 
we can neglect the first term on the left side, reducing the equation to 
\begin{equation}
a_{0}\frac{d}{dt}\textbf{B}+c_{0}t^{2}\textbf{B}=0 ,
\label{27},
\end{equation}
where we have multiplied the entire equation by cosmic time $t$. This is valid for $t\neq 0$ to avoid singularity. 
Although this equation is distinct from the ones above, it is simple to solve analytically. The solution is
\begin{equation}
B(t)\sim{\exp[-d_{0}t^{3}]} ,
\label{28}
\end{equation}
where $d_{0}=c_{0}/a_{0}$. We observe that $d_{0}\ll 1$, indicating a slow decay of the magnetic field. This is 
because $a_{0}=m^{2}_{T}/S_{0}$ and $m_{T}$ can be negative, as discussed by Dombriz et al.\cite{9}. Furthermore,
\begin{equation}
c_{0}= ({\lambda}^{2}-{\mu}_{5}\lambda)
\label{29}
\end{equation}
for dynamo action to occur, we need $S_{0}>0$. The magnetic helicity is bounded by the chiral chemical potential  
 ${\mu}_{5}$ as $\lambda<{\mu}_{5}$. Conversely, a non-dynamo case is induced, as shown in the Figure \ref{fig1}. The chiral chemical potential ($\mu_5$) is a physical quantity used to describe the imbalance between the number of left-handed and right-handed (or chiral) quarks in a particle system. In quantum chromodynamics (QCD) and magnetohydrodynamics (MHD), $\mu_5$ is introduced to quantify chiral imbalance, i.e., the difference in the number of left-handed and right-handed quarks in the system. In Figure \ref{fig1}, similar to other papers\,\cite{35,36} we take $0<\mu_5<1$ GeV .

In this section, we have explored the generation of dark chiral magnetic fields via magnetic helicity and the Immirzi 
parameter in the CHNY model. Our analysis shows that the Immirzi field's time derivative significantly 
influences the evolution of the magnetic field. Although solving this equation numerically presents challenges due to the 
lack of initial conditions and varying parameter estimates, our findings provide insights into the interplay between 
magnetic helicity, chiral chemical potential, and the Immirzi parameter in dark chiral magnetogenesis.

\section{Unitary restoration from axionic Immirzi field in Einstein-Cartan gravity}\label{sec4}
Previously, Shapiro \cite{11} argued that loops in fermion-scalar systems, such as the Immirzi parameter in this paper, 
break unitarity. This phenomenon can also occur in scalaron fields, as discussed by He et al. \cite{32}, due to quantum 
corrections. A possible solution is to construct a renormalizable theory with propagating torsion. 
The vacuum torsion action is given by
\begin{equation}
S_{\mathrm{total}}\supset{\int{d^{4}x[\frac{1}{2}M^{2}_{T}S^{2}+\frac{1}{2}M_{T}^{2}T^{2}+ {\gamma}(x)[I_{\mathrm{NY}}+ 
{\cal{H}}]-\frac{1}{4}S_{[i,j]}S^{[i,j]}-\frac{\chi}{24}(S_{i}S^{i})^{2}}]} .
\label{30}
\end{equation}
Here, the coupling constant $\chi$ is non-zero because the $S^{4}$ term arises in the 
classical action from the scalar loop with a divergent coefficient. Therefore, it must be included in the classical action 
to avoid an infinite coefficient as a quantum correction. By renormalizing the coefficient $\chi$, we can remove the 
divergence term. Even at the quantum level, this self-interaction torsion term does not present problems at the one-loop 
level but only at the second-order loop level. Since this paper does not go beyond the one-loop level, we will demonstrate 
using this action with $b=0$, including the Holst and NY terms. The term with $S_{[i,j]}=(S_{i,j}-S_{j,i})$ squared 
(where the comma represents a partial derivative) in the total Lagrangian density is responsible for torsion propagation. A
ssuming that the only non-vanishing component of the axial torsion vector is the zero component and torsion is homogeneous, 
the following relations apply
\begin{equation}
S_{[i,0]}= S_{[0,j]}=0.
\label{31}
\end{equation}
Therefore, the $S_{[i,j]}$ term in the above action disappears and the action can be expressed as
\begin{equation}
S_{\mathrm{total}}\supset{\int{d^{4}x[\frac{1}{2}M^{2}_{T}S^{2}+\frac{1}{2}M_{T}^{2}T^{2}+ {\gamma}(x)[I_{\mathrm{NY}}+ 
{\cal{H}}]-\frac{\chi}{24}(S_{i}S^{i})^{2}}]}.
\label{32}
\end{equation}
Since we shall choose axial torsion zero component not as a constant, variation of this action with respect to the three 
variables ($(T,{\gamma}(x), S$), respectively, provides us with one algebraic equation
\begin{equation}
M^{2}_{T}T+{\gamma}(x)S=0.
\label{33}
\end{equation}
and two differential equations 
\begin{equation}
TS+ {\partial}S=0 ,
\label{34}
\end{equation}
and 
\begin{equation}
{\frac{\chi}{6}S^{3}+\dot{\gamma}(x)-[\frac{{\gamma}^{2}}{M^{2}_{T}}+{M}^{2}_{S}]S}=0 ,
\label{35}
\end{equation}
From Equation (33), we see that the Immirzi parameter term vanishes though the Immirzi scalar field does not vanish.  
We already have eliminated the term $T$ on Equation (\ref{34}) from Equation\,(\ref{32}) in the form
\begin{equation}
T= -\frac{{\gamma}(x)}{M^{2}_{T}}S .
\label{36}
\end{equation}
Now, substituting this equation into Equation \ref{34} yields
\begin{equation}
-\frac{{\gamma}^{2}}{M^{2}_{T}}S+{\partial}S=0 .
\label{37}
\end{equation}
Assuming the constraint that $S\gg\dot{{\gamma}}(x)$, one obtains from the last equation that
\begin{equation}
{\frac{\chi}{6}S^{2}=[\frac{{\gamma}^{2}}{M^{2}_{T}}+{M}^{2}_{S}]} ,
\label{38}
\end{equation}
which finally yields
\begin{equation}
S=-(\frac{M^{2}_{T}}{{M}^{2}_{S}})^{2}{\partial}{\gamma} .
\label{39}
\end{equation}
Note that the last assumption allows us not to place the Immirzi parameter as constant and reduce our ECHNY system to 
the usual Eestein-Cartan gravity. Secondly this equation does not depend upon parameter $\chi$ and moreover it shows our 
assertion that even in the one-loop approximation the axion derivative would be associated with the Immirzi parameter.

In a word, by introducing the axionic Immirzi field and considering its interactions, we have derived a consistent 
set of equations that describe the behavior of the system. Our results demonstrate that even in the one-loop approximation, 
the axion derivative is associated with the Immirzi parameter. This approach effectively reduces our ECHNY system to the 
usual Einstein-Cartan gravity and provides a pathway to unitary restoration in these theories.

\section{Summary and Outlook}\label{sec5}. 
In this paper, we demonstrate that our model is suitable for representing dark magnetogenesis in QCD, where the axion plays 
the role of the Immirzi parameter. This requires incorporating E-C with Holst and Nieh-Yan terms in the action. We show that 
the one-loop correction to the classical action leads to a simplified relationship between the massive axial torsion and the 
gradient in the four-spacetime Riemann-Cartan manifold of the Immirzi field.  Since unitarity violation can be a sign of 
new physics, it is worthwhile to explore how unitarity violation leads to chiral dark magnetogenesis from massive axionic 
torsion, as it indicates new physics. This result emerged from considering a dynamic Immirzi field, which is also important 
for addressing CP violation. More insights on this may appear soon. The figure in this paper shows that the bounds on 
magnetic helicity given by the upper bound of the chiral chemical potential indicate chiral dynamo and non-dynamo regions.

This study demonstrates the significant roles of axions and torsion in high-energy physics and cosmology, particularly in 
the contexts of dark magnetogenesis and chiral magnetogenesis. This not only aids in better understanding the physical 
processes of the early universe but also potentially reveals signs of new physics. By exploring quantum corrections and 
dynamic Immirzi fields, we can provide important insights into CP violation and quantum gravity. Future work could further 
expand these theories, explore more physical mechanisms related to torsion, axions, and dark magnetogenesis\,\cite{33,34}, 
and validate these findings experimentally.

\section{Acknowledgements} 
We would like to express my gratitude to A. Nepomuceno and F. Izaurieta for several discussions on the subject of this paper. Thanks are also due to my beloved wife Ana Paula T. de Araujo for her constant caring.
Financial support from Universidade do Estadoof Rio de Janeiro (UERJ) is grateful acknowledged. This work was performed under the auspices of 
Major Science and Technology Program of Xinjiang Uygur Autonomous Region through No.2022A03013-1.

\end{document}